\documentclass[letterpaper]{article}
\usepackage{aaai}
\usepackage{times}
\usepackage{helvet}
\usepackage{courier}

\usepackage{pstricks}
\usepackage{pst-node}
\usepackage{latexsym}
\usepackage{graphicx}
\usepackage{amsmath}
\usepackage{amsfonts}
\usepackage{amssymb}
\usepackage{epstopdf}
\usepackage{algorithm}
\usepackage[noend]{algorithmic} 
\usepackage{amsthm}

\newtheorem{proposition}{Proposition}

\newtheorem{example}{Example}

\begin{document}

\title{Mechanism Design for Mobile Geo--Location Advertising }
\author{Nicola Gatti \and Marco Rocco \\
Politecnico di Milano \\
Piazza Leonardo da Vinci 32 \\
Milano, Italy \\
\{nicola.gatti, marco.rocco\}@polimi.it\\
\And
Sofia Ceppi \\
Microsoft Research \\
21 Station Road \\
Cambridge, CB1 2FB, UK \\
soceppi@microsoft.com\\
\And
Enrico H. Gerding \\
University of Southampton\\
University Road, Highfield\\
Southampton, SO17 1BJ, UK\\
eg@ecs.soton.ac.uk\\
}
\maketitle
\begin{abstract}
%\vspace{-0.2cm}
Mobile geo--location advertising, where mobile ads are targeted based on a user's location, has been identified as a key growth factor for the mobile market. As with online advertising, a crucial ingredient for their success is the development of effective economic mechanisms. An important difference is that mobile ads are shown sequentially  over time and information about the user can be learned based on their movements. Furthermore, ads need to be shown selectively to prevent ad fatigue. To this end, we introduce, for the first time, a user model and suitable economic mechanisms which take these factors into account. Specifically, we design two truthful mechanisms which produce an advertisement \emph{plan} based on the user's movements. One mechanism is allocatively efficient, but requires exponential compute time in the worst case. The other requires polynomial time, but  is not allocatively efficient. Finally, we experimentally evaluate the trade--off between compute time and efficiency of our mechanisms.
\end{abstract}

%\vspace{-0.15cm}
\section{Introduction}
\label{introduction}
%\vspace{-0.05cm}
\noindent Mobile geo--location advertising~\cite{Vallina-Rodriguez:2012:BCC:2398776.2398812}, where mobile ads are targeted based on a user's location (e.g., streets or squares), has been identified as a key growth factor for the mobile market. Growing at an annual growth rate of 31\%, the mobile ad market is forecasted to be worth 19.7 billion Euros in 2017---about 15.5\% of the total digital advertising market~\cite{Berg}. A crucial ingredient for its success will be the development of effective economic mechanisms. 
%However, to date,  problem has not been addressed in the literature. 
To this end, we propose, for the first time, economic mechanisms for the mobile geo--location advertising scenario addressing three issues: modeling the \emph{users' behaviour}, avoiding \emph{advertisers' strategic manipulation}, and designing \emph{tractable algorithms}.

%To date, much of the literature on computational advertising has focused on sponsored search, where ads are shown alongside search results. This type of advertising is based on so--called \emph{sponsored search auctions}~\cite{Narahari2009}, in which advertisers bid for keywords and they pay only if their links are clicked.
To date, much of the literature on computational advertising has focused on \emph{sponsored search auctions}~\cite{Narahari2009}, in which advertisers bid for keywords and pay only if their links are clicked.
A crucial ingredient for their success is the allocation and payment mechanism.
%The most widely used auction is the  Generalized Second Price (GSP), which allocates ads based a combination of the bids and ad quality (which is derived from the historic click through rate), and computes the payment based on the ad shown in the next slot.
The most widely used auction is the Generalized Second Price (GSP) but another well--known mechanism, the Vickrey--Clarke--Groves (VCG) mechanism~\cite{Narahari2009}, is being applied in websites such as Facebook. An important research issue for analysing such auctions is the modeling of  user behaviour, where it is typically assumed that users scan displayed ads from the top to the bottom in a Markovian fashion~\cite{DBLP:conf/wine/AggarwalFMP08,cascade}.
%Fotakis,
%,Craswell,Joachims
 Such a user model induces so-called {\it externalities}, where the click probability of an ad further down the list depends on the ads which are shown earlier on.  More recently, there is also increasing research on display ads, e.g. banners on websites, where advertisers are matched to publishers (webpages) through a complex web of ad networks and exchanges~\cite{muthukrishnan2010ad}. 

% webpages search has moved to %Machine learning techniques are then exploited to estimate the user's model parameters~\cite{lazaric}.
 %Although it has been is not \emph{incentive compatible}~\cite{edelman} and, when, as usual, information is uncertain, the GSP can admit Bayes--Nash equilibria that are much worse (up to $\frac{1}{8}$) w.r.t. the Vickery--Clarke--Groves (VCG) truthful equilibrium~\cite{paes}. For this reason, recently, some websites adopt the VCG, e.g. Facebook.
%\vspace{-0.05cm}
%However, none of these online advertising models can be directly applied to mobile geo--location, e.g., due to a limited screen size, which means that ads cannot be shown simultaneously.
However, none of these online advertising models can be directly applied to mobile geo--location mainly because they do not take into account the future behaviour of the user.
Rather, mobile ads, such as coupons and ads in mobile apps, are shown sequentially over time while the user moves in an environment, e.g. a city or shopping centre. Furthermore, users are affected by the same ad in different ways depending on the location in which they receive the ads (e.g., if the shop is far from the location in which the ad is received, users are more likely to discard the ad), and the path followed so far can reveal information about the user's intention (i.e., the user's next visits). Thus, ad allocations can be done dynamically taking the user behaviour into account, unlike in sponsored search auctions, where the entire allocation is shown simultaneously. 

To address these problems, we design the first model for mobile geo--location advertising, which calculates an advertising plan based on the path followed so far and predicted future path. We adopt a \emph{pay--per--visit} scheme, where an advertiser pays only if a user actually visits the shop after having received the ad (based on geo--location or by redeeming a coupon). Importantly, we consider user models where the visit probability depends on their position. In addition, we capture the ad fatigue phenomenon~\cite{Abrams:2007:PAD:1781894.1781964}, discounting the visit probability associated with the  next ads as a user receives more ads. This creates sponsored--search like externalities, except that the visit probability depends only on the number of ads shown prior, and not on which ads are shown. Then, we focus on the problems of developing novel allocation algorithms, both optimal and approximate, for designing incentive compatible mechanisms. We analyse theoretical bounds on their performance and experimentally evaluate them. %For reasons of space, we report proofs and examples in the supplemental material~\cite{supplementalmaterial}.
%online appendix~\cite{appendix}.
%
%\begin{center}
\vspace{-0.35cm}
\begin{figure}[!h]
\includegraphics[scale=0.83]{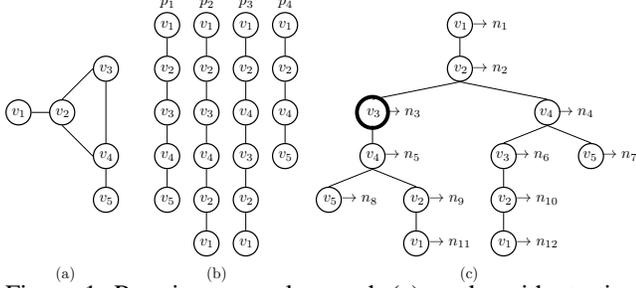}
\vspace{-0.85cm}
\caption{Running example: graph (a), paths with starting vertex $v_1$ (b), tree of paths (c).}\label{fig::total}
\end{figure}
%\vspace{-0.4cm} 
%\end{center}
%\vspace{-0.05cm}
%
%
%
%\vspace{-0.3cm}
\section{Problem Statement}
\label{sec:problemstatement}
%\vspace{-0.1cm}
\textbf{User mobility model}. We represent a physical area, e.g. a city, as a graph $G = \{V,E\}$, e.g. Fig.~\ref{fig::total}(a), where $V$ is the set of vertices~$v$ and $E$ is the set of edges. Vertices are subareas, e.g. streets or squares, in which an ad can be sent to a user. A user will move over the graph following a path, denoted by $p \in P$ and defined as a sequence of adjacent vertices, where $P$ is the set of possible paths. We denote the first or \emph{starting vertex} of a path $p$ by $v_s(p)$, and partition paths by this starting position. To this end, we introduce $P_v \subseteq P$ which denotes the set of paths $p$ with $v_s(p)=v$. For instance, Fig.~\ref{fig::total}(b) depicts 4 paths with starting vertex $v_1$. In addition, we associate each path $p$ with a probability $\gamma_p$---estimated, e.g., by means of machine learning tools~\cite{lazaric}---that indicates how likely the given path will be followed by users, given the starting vertex of the user.   Thus, $\sum_{p\in P_v}\gamma_p=1$. Since the number of possible paths in $P_v$ can be arbitrarily large, we restrict $P_v$ to a finite (given) number of paths containing only the paths with the highest probability. We normalize the probabilities $\gamma_p$ accordingly. 

Given a user's actual starting vertex, we can build the tree of the paths the user could follow. Fig.~\ref{fig::total}(c) depicts an example of such a tree with starting vertex $v_1$.
%\footnote{\scriptsize We notice that an alternative user mobility model based on Markov chains would not be satisfactory, because users are not Markovian.} 
We denote by $N = \{n_1,\ldots, n_{|N|}\}$ the set of tree nodes where each node $n$ is associated with a single graph vertex $v \in V$, whereas each vertex $v$ can be associated with multiple tree nodes $n \in N$. We define $\alpha_n$ as the probability with which node $n$ is visited by a user given a starting vertex. In particular, $\alpha_n$ is equal to the sum of the $\gamma_{p}$ of the paths $p$ sharing $n$. Consider, for instance, the bold node $n_3$ in Fig.~\ref{fig::total}(c): since $n_3$ is contained in both the paths $p_1$ and $p_2$, $\alpha_{n_3} = \gamma_{p_1} + \gamma_{p_2}$.

\textbf{Advertising model}. Let $A = \{a_1,\ldots,a_{|A|}\}$ denote the set of advertisers. W.l.o.g., we assume each advertiser to have a single ad (thus we identify both advertiser and ad by $a$). We denote by $\theta$ an \emph{advertising plan} (i.e., an allocation of ads to nodes in which the ads are sent to the users) and by $\Theta$ the set of all the advertising plans. Formally, $\theta$ is a function $\theta:N\rightarrow A\cup\{a_\emptyset\}$ mapping each node of a tree of paths to an ad, where $a_\emptyset$ is an empty ad corresponding to sending no ad. We constrain $\theta$ to not allocate the same ad (except $a_\emptyset$) on different nodes belonging to the same path 
(the basic idea is that receiving the same ad multiple times does not affect the visit probability of the corresponding shop, but it just increases ad fatigue),
%(the basic idea is that the behaviour of a user would not be affected by receiving the same ad multiple times and thus allocating the same ad more than once introduces a negative externality)
while the same ad can be allocated on different paths. Each advertiser receives a reward $r_{a} \in R\subseteq \mathbb{R}^+$ from the visit of a user to his shop; we define $r_{a_\emptyset}=0$ and ${\bf r} = (r_{a_1}, \ldots, r_{a_{|N|}})$ the profile of rewards. The probability with which a user visits the shop $a$ is given by $VTR_a(\theta)$, also called \emph{visit through rate}, which is determined by the user attention model as specified below. Given this, the expected reward of $a$ from an advertising plan $\theta$ is:
%
%\vspace{-0.3cm}
%\begin{scriptsize}
%\[
%\mathbb{E}[r_a|\theta]=VTR_a(\theta) \cdot r_{a},
%\]
%\end{scriptsize}
%\vspace{-0.4cm}
%
%\noindent and we define the social welfare $SW$ over $\theta$ by:
%
%\vspace{-0.3cm}
%\begin{scriptsize}
%\[
%SW(\theta)=\sum_{a\in A}\mathbb{E}[r_a|\theta]
%\]
%\end{scriptsize}
%\vspace{-0.3cm}
%
%
%\begin{scriptsize} $\mathbb{E}[r_a|\theta]=VTR_a(\theta) \cdot r_{a}$\end{scriptsize}, and we define the social welfare $SW$ over $\theta$ by: \begin{scriptsize} $SW(\theta)=\sum_{a\in A}\mathbb{E}[r_a|\theta]$\end{scriptsize}.
$\mathbb{E}[r_a|\theta]=VTR_a(\theta) \cdot r_{a}$, and we define the social welfare $SW$ over $\theta$ by: $SW(\theta)=\sum_{a\in A}\mathbb{E}[r_a|\theta]$.

\textbf{User attention model.}
We assume that the VTR of an ad $a$ depends on both \emph{where} the ad is shown on the path (i.e., depending on whether the ad is relevant to the current position) and {\it number of other ads} shown prior to this one (due to \emph{ad fatigue}). To this end, we define function $c:\Theta\times N\rightarrow \mathbb{N}$ returning the number of \emph{non--empty} ads allocated to nodes that precede $n$ from the root node. The VTR of an ad $a \in A$ given advertising plan $\theta$ is: 
%
%\vspace{-0.3cm}
%\begin{scriptsize}
%\[
%VTR_a(\theta)= \sum_{n\in N:\theta(n)=a} \alpha_{n}\cdot \Lambda_{c(\theta,n)}\cdot q_{a,n},
%\]
%\end{scriptsize}
%\vspace{-0.3cm}
%
%\noindent where:
$VTR_a(\theta)= \sum_{n\in N:\theta(n)=a} \alpha_{n}\cdot \Lambda_{c(\theta,n)}\cdot q_{a,n}$, where:
\vspace{-0.14cm}
\begin{itemize}
\item $\alpha_{n}$ is the probability that node $n$ is visited;
\vspace{-0.14cm}
\item $\Lambda_{c(\theta,n)}$ is the \emph{aggregated continuation probability}, capturing the ad fatigue phenomenon where the user's attention decreases as more ads are received; we assume the user attention decreases as $\Lambda_{c(\theta,n)}= \prod_{i=1}^{c(\theta,n)} \lambda_i$ where $\lambda_i \in [0,1] \ \forall i \in \{1,\ldots, |N|\}$ and $\Lambda_0=1$;
\vspace{-0.14cm}
\item $q_{a,n}\in[0,1]$ is the relevance or \emph{quality} of ad $a$ at node $n$, representing the VTR when $a$ is the first allocated ad in all the paths.
\end{itemize}
\vspace{-0.15cm}

\textbf{Mechanism design problem}. Our aim is to design a direct--revelation economic mechanism $\mathcal{M}=(A,\Theta,R,f,t)$, where the agents $A$ are the advertisers, the outcomes $\Theta$ are the possible advertising plans, the agents communicate their (potentially untruthful) valuation over the allocations reporting their single--parameter reward $\hat r_a$. Note that, similar to sponsored search, we assume the ad qualities to be known by the system and not part of the report. The agents' valuation space is $R$, the allocation function $f:R^{|A|}\rightarrow \Theta$ maps the profile $(\hat{r}_{a_1},\ldots,\hat{r}_{a_{|A|}})$ constituted of the valuations reported by the agents to the space of the possible advertising plans, and the transfer function $t_a:R^{|A|}\rightarrow \mathbb{R}$ maps reported valuations profiles to the monetary transfer of each agent~$a$. The aim of each advertiser is to maximise his own utility $u_a(\hat{\bf{r}}, r_a) = VTR_a(f(\hat{\bf{r}})) r_a - t_a(\hat{\bf{r}})$. This is the reason an advertiser could be interested in misreporting his true reward ($\hat{r}_a \not = r_a$). We aim at designing $f$ and $t_a$ such that dominant strategy incentive compatibility (DSIC), individual rationality (IR), weak budget balance (WBB), and allocative efficiency (AE) \cite{Narahari2009} are satisfied. Moreover, we would like $f$ and $t$ to be efficiently computable (polynomial time).

%\vspace{-0.2cm}
\section{Single--Path Case}%
\label{sec:singlepath}
%\vspace{-0.05cm}
\noindent In this section, we study the basic case with a single path, before proceeding to the more general case. %We first present a mechanism with an efficient allocation function $f_E$, and then go on to present a mechanism with an approximate allocation function $f_A$. 
%\vspace{-0.25cm}
\subsection{Efficient Mechanism}
%\vspace{-0.1cm}
%\textbf{Mathematical programming formulation for the allocation function}. 
\noindent The main challenge of an efficient mechanism is finding the optimal advertising plan, i.e. the allocation of ads to nodes that maximises social welfare, subject to the constraint that each ad $a \in A$ can appear at most once on the (single) path (except ad $a_\emptyset$). To this end, we show that this allocation problem is a variation of a known \emph{linear assignment problem} (AP) as in Burkard \emph{et al.}~\shortcite{assignment}. In particular, when the aggregated continuation probability, $\Lambda_c$, is a constant (i.e., when $\lambda_i=1\ \forall i$, which means that there are no externalities) the single--path problem corresponds to the classical 2--index AP (2AP), where the aim is to allocate a set of tasks to a set of agents while minimising/maximising the sum of costs/profits $w$, subject to each agent having exactly one task. In our problem, agents correspond to nodes $N$, and tasks to ads $A$. Furthermore, since there is a single path, we have that $\alpha_n=1, \forall n \in N$. Then, for $\lambda_i=1\ \forall i$, the \emph{value} for an ad--node assignment $(a,n)$ is given by the expected reward $E[r_a|\theta(n)=a] = w_{a,n} = \hat{r}_{a}\cdot q_{a,n}$ (to maximise). 

Now, it is well known that the 2AP can be solved in polynomial time by means of the Hungarian algorithm---with complexity $O(\max\{|N|^3,|A|^3\})$---or linear programming (LP) as in Burkard \emph{et al.}~\shortcite{assignment}, where the continuous relaxation results in a basic integer solution. However, when $\lambda_i<1$, the nature of our problem becomes fundamentally different. In particular, the 2AP optimal solution always requires that all the agents are assigned with a task. This also holds in our setting when $\lambda_i=1 \ \forall i$.
However, when $\lambda_i<1$, it can be optimal to leave some nodes unallocated, as shown in the following example:
%However, when $\lambda_i<1$, it can be optimal to leave some nodes unallocated as shown in the supplemental material~\cite{supplementalmaterial}.
%\vspace{-0.2cm}

\begin{table}[t]
\begin{scriptsize}
\centering
\begin{tabular}{|c||c|c|c|c||c|c|}\cline{1-3} \cline{5-7} 
		& $\hat{r}_a \cdot q_{a,n_1}$ & $\hat{r}_a \cdot q_{a,n_2}$ &  &  & $\hat{r}'_a \cdot q_{a,n_1}$ & $\hat{r}'_a \cdot q_{a,n_2}$ \\ \cline{1-3} \cline{5-7} \cline{1-3} \cline{5-7} 
$a_1$	& \textbf{1} & 2 &  &  $a_1$ & 3 & \textbf{6}\\ \cline{1-3} \cline{5-7}  
$a_2$	& 2 & \textbf{4} &  &  $a_2$ & \textbf{2} & 4\\ \cline{1-3} \cline{5-7} 
\multicolumn{7}{c}{}\vspace{-0.3cm}\\ 
\multicolumn{3}{c}{(a)} & \multicolumn{1}{c}{} & \multicolumn{3}{c}{(b)} 
\end{tabular}
\end{scriptsize}
%\vspace{-0.3cm}
\caption{Two scenarios used as examples.}
\label{table::LTAP}
%\vspace{-0.7cm}
\end{table}

\begin{example}
Consider the case represented in Table~\ref{table::LTAP}(a) where $N = \{n_1,n_2\}$, $A=\{a_1,a_2\}$. The optimal solution of the 2--index AP allocates $a_1$ in $n_1$, and $a_2$ in $n_2$ (in bold in the table). Instead, with $\lambda_1  < 0.75$, the optimal solution of our problem allocates only $a_2$ in $n_2$.
%\vspace{-0.1cm}
\end{example}

When $\lambda_i<1$, our allocation problem can be formulated as a variation of the \emph{3--index assignment problem} (3AP) as:
%
%\vspace{-0.2cm}
\begin{scriptsize}
\[ 
\max_{\theta \in \Theta} \sum_{a \in A}\sum_{n \in N} \sum_{c \in C} \Lambda_c \cdot \hat{r}_{a} \cdot q_{a,n} \cdot x_{a,n,c} 
\]
\vspace{-0.4cm}
\begin{align}
\sum_{n \in N} \sum_{c \in C} x_{a,n,c} \leq 1 & \quad \forall a \in A \label{eq::onet}\\
\sum_{a \in A} \sum_{c \in C} x_{a,n,c} \leq 1 & \quad \forall n \in N \label{eq::oneslot}\\
\sum_{a \in A} \sum_{n \in N} x_{a,n,c} \leq  1 & \quad \forall c \in C \label{eq::oneposition}\\
\sum_{a \in A} x_{a,n,c} - \sum_{a \in A} \sum_{\underset{n' < n}{n' \in N:}} x_{a,n',c-1} \leq 0& \quad \forall n \in N, c \in C \setminus \{0\} \label{eq::ordered} \vspace{-0.1cm}\\
x_{a,n,c} \in \{0,1\} &  \quad  \forall a \in A, n \in N, c \in C 
\end{align}
\end{scriptsize}
\vspace{-0.5cm}

\noindent where $x_{a,n,c}=1$ if $a=\theta(n)$ (i.e. ad~$a$ is allocated to node~$n$) and $c=c(\theta,n)$ (i.e. $a$ is the $c+1$--th allocated non--empty ad along the path); $x_{a,n,c}=0$ otherwise. $C=\{0,\ldots, |N|-1\}$ contains all the possible values of~$c$. Constraints~(\ref{eq::onet}) ensure that each ad $a \neq a_{\emptyset}$ is allocated at most once;  Constraints~(\ref{eq::oneslot}) ensure that each node is allocated to an ad $a \neq a_{\emptyset}$ at most once; Constraints~(\ref{eq::oneposition}) ensure that there cannot be two ads with the same number of preceding ads (except for the empty ad); Constraints~(\ref{eq::ordered}) ensure that, whenever $x_{a,n,c}=1$ (i.e. if some ad $a \neq a_{\emptyset}$ is allocated to a node~$n$ with $c$ preceding  ads), then $c$ ads must be actually allocated in the path preceding $n$.

\textbf{Complexity issues}. Compared to the 3AP formulation, our problem has Constraints~(\ref{eq::ordered}) as additional constraints. Moreover, our objective function is a special case of the 3AP  objective function (the original  3AP function is given by $\max \sum_{a \in A}\sum_{n \in N} \sum_{c \in C} w_{a,n,c} \cdot x_{a,n,c} $). The maximization version of 3AP is easily shown to be $\mathcal{NP}$--hard by reduction from the \emph{3--dimentional matching problem} (3DMP), however, there is no straightforward reduction from $\mathcal{NP}$--hard problems to ours.
%Furthermore, we can show that the continuous relaxation of our allocation problem admits, differently from 2AP, non--integer optimal solutions and thus the above integer mathematical programming formulation cannot be solved (in polynomial time) by LP tools, as shown into the supplemental material~\cite{supplementalmaterial}.

Furthermore, we can show that the continuous relaxation of our allocation problem admits, differently from 2--index AP, non--integer optimal solutions and thus the above integer mathematical programming formulation cannot be solved (in polynomial time) by LP tools, as shown by the following example.
%\vspace{-0.2cm}
\begin{example}
We have $A = \{a_1,a_2,a_3 \}$, $N = \{n_1, n_2, n_3\}$, and $\lambda_i = 0.2\ \forall i$. Parameters $q_{a,n}$ and $\hat{r}_a$ are: $q_{a_1,n_3}=q_{a_2,n_1}=q_{a_3,n_2}=1$, while all the others $q_{a,n}=0$, and $\hat{r}_{a_1}=100$, $\hat{r}_{a_2}=79$, $\hat{r}_{a_3}=70$.
\begin{table}[!h]
\vspace{-0.2cm}
\label{ex::noneff}
\begin{center}

\begin{scriptsize}
\begin{tabular}{cc}
\begin{tabular}{|c|c|c|c|}
\hline
		&	$q_{a,n_1}$	&	$q_{a,n_2}$	&	$q_{a,n_3}$	\\
\hline
$a_1$		&	$0$	&	$0$	&	$1$	\\
\hline
$a_2$		&	$1$	&	$0$	&	$0$	\\
\hline
$a_3$		&	$0$	&	$1$	&	$0$	\\
\hline
\end{tabular}

&

\begin{tabular}{|c|c|}
\hline
			&	$r_a$	\\
\hline
$a_1$		&	$100$	\\
\hline
$a_2$		&	$79$	\\
\hline
$a_3$		&	$70$	\\
\hline
\end{tabular}

\end{tabular}
\end{scriptsize}
\end{center}
%\vspace{-0.2cm}
\end{table}

\vspace{-0.3cm}
\noindent The optimal solution of the continuous relaxation of our allocation problem is: $x_{a_1, n_3, c_0} = x_{a_1, n_3, c_1} = x_{a_2, n_1, c_0} = x_{a_3, n_2, c_1} = 0.5$, with a social welfare of $106.5$. Instead, the optimal integer solution is:  $x_{a_1, n_3, c_0} = 1$, with a social welfare of $100$.
\end{example}

%Therefore, the theoretical hardness result for our allocation setting is still an open problem and we leave this for future work. Instead, in this paper we focus on the problem of designing an algorithm for finding the optimal solution, and identifying special cases for which computing the solution is a polynomial time task. We then proceed, in Section~\ref{sec:approx}, with presenting approximate solutions.

We leave as future work a more accurate theoretical study of the hardness of the problem, instead, this paper focus on the problem of designing an algorithm for finding the optimal solution, and identifying special cases for which computing the solution is a polynomial time task. We then proceed, below, with presenting approximate solutions.

\textbf{Allocation function algorithm in unrestricted domains}. We start by considering the unrestricted setting. For this setting, any branch--and--bound algorithm enumerating all the allocations, e.g., using standard integer programming or~\cite{balas1991} for 3AP, has a complexity of $O(|A|^{|N|})$ in the worst case. We show that it is possible to have an algorithm for $f_E$ with a better complexity.

Our algorithm, named \textsf{OptimalSinglePath}, works as follows. First, we split the problem into subproblems. In detail, let $B \subseteq N$ denote a set of nodes such that we assign non--empty ads ($a \not= a_{\emptyset}$) to all nodes $n \in B$, and empty ads ($a_{\emptyset}$) to all nodes $n \notin B$. Note that there are exactly $2^{|N|}$ such combinations (assuming for the sake of simplicity that $|A|\geq|N|$). Now, for a given combination $B$, the number of nodes with non--empty ads preceding any $n \in B$ is fixed. Let this number be denoted by $c(B,n)$. Then, the problem of finding the optimal allocation for a given $B$ can be formulated as an AP where $w_{a,n}=\hat{r}_a\cdot q_{a,n}\cdot \Lambda_{c(B,n)}$. Therefore, it can be solved by using an AP--solving oracle with a complexity of  $O(|A|^3)$. Our algorithm then calls the AP--solving oracle for each $B \subseteq N$. Finally, the algorithm returns the best found allocation. The complexity is $O(2^{|N|}\cdot |A|^3)$.

\textbf{Allocation function algorithm in restricted domains}. We consider two restricted domains in which $f_E$ is easy.

\emph{Node--independent qualities}. Assume that, for every ad $a \in A$, the following holds: $q_{a,n} = q_{a,n'} = q_a$ for all $n, n' \in N$. In words, the visit probability does not depend on the specific node where the ad is shown, but only on the ad itself, and the number of preceding ads shown. In this case, the mobile geo--location advertising reduces to the sponsored search auctions with only position--dependent externalities that is known to be easy~\cite{cascade}.
%The algorithm for the allocation function $f_E$ sorts the ads in decreasing order of $q_{a} \cdot \hat{r}_{a}$ and then allocates the ads to the slot accordingly.
Notice that, in this special case, the optimal advertising plan prescribes that all the slots (nodes) are filled with an ad.

\emph{Single--node maximal ads}. We say that ad $a$ is \emph{maximal} for a given node $n$, denoted by $a^{\max}_n$, if $a$ is the best ad (in terms of expected value) for node $n$. Formally: $a_n^{\max} = \arg \max\limits_{a \in A} \{q_{a,n} \hat{r}_a\}$. Assume that each ad is maximal in at most a single node of the path. Formally: $a_n^{\max} \not = a_{n'}^{\max}$ for all $n, n' \in N$ with $n \not = n'$. This is reasonable when there are many ads and the quality strongly depends on the distance between the shop and the current position of the user, e.g., the user decides to visit the shop only if it is right next to him. In this case, if the algorithm allocates an ad to a given node, then it will allocate the maximal ad (this is not the case if an ad is maximal in multiple nodes).
%Therefore, at each node, all the non--maximal ads can be discarded. In what follows, we describe an algorithm for this setting based on dynamic programming that divides the main problem into subproblems, and progressively builds the optimal solution in polynomial time.

%Our algorithm, summarized in Algorithm~\ref{alg:maximal}, works as follows. 
The algorithm we proposed (Algorithm~\ref{alg:maximal}), based on dynamic programming, works as follows.
First, suppose that nodes of set $N$ are numbered in increasing order from the root $n_1$ to the leaf $n_{|N|}$. Each subproblem is characterized by a pair $[i,j]$ with $i\in \{0, \ldots, |N|-1\}$ and $j\in\{1,\ldots,|N|\}$ and aims at finding the optimal allocation of the subpath of nodes from $n_j$ to $n_{|N|}$ when the number of ads allocated in the subpath of nodes from $n_1$ to $n_{j-1}$ is $i$. The rationale of the algorithm is to start from the leaf of the path and to move backward given that, in the case each node has a different maximal, the optimal allocation of a subproblem $[i,j]$ does not depend on the optimal allocation of a subproblem $[i',j']$ strictly including $[i,j]$, i.e., $i'\leq i$ and $j'<j$. 

We use two $|N| \times |N|$ matrices $\Pi$ and $\Phi$. Each element $\Phi[i,j]$ is the optimal allocation of subproblem $[i,j]$ and it is represented as a set of pairs $(a,n)$ where $a$ is the ad allocated in $n$; while each element $\Pi[i,j]$ is  the expected value of the optimal allocation of subproblem $[i,j]$. %In Algorithm~\ref{alg:maximal} we write $i \in \langle \alpha, \ldots, \beta\rangle$ to indicate that variable $i$ assumes values in the same sequence they are in the ordered set $\langle \alpha, \ldots, \beta\rangle$.

At Steps~(\ref{st:lastclbeg}--\ref{st:lastclend}) the algorithm fills all the elements of the last column of $\Pi$, i.e. $\Pi[i,|N|]$ $\forall i \in \{0, \ldots, |N|-1\}$, with the value $\Lambda_{i} \cdot q_{a^{\max}_{n_{|N|}}, n_{|N|}} \cdot \hat{r}_{a^{\max}_{n_{|N|}}}$, i.e. the contribute that ad $a^{\max}_{n_{|N|}}$ provides to the social welfare when $a^{\max}_{n_{|N|}}$ is allocated in node $n_{|N|}$ when it is the $i+1$--th allocated ad. Indeed, any optimal allocation will have an ad allocated in the last node. Then, at Steps~(\ref{st:otherbeg}--\ref{st:otherend}), the algorithm selects each node $n_j$ from $n_{{|N|}-1}$ to $n_1$, and finds the optimal advertising plan for the subpath from $n_j$ to $n_{|N|}$. Consider a generic element $\Pi[i,j]$, the algorithm decides whether it is better to allocate $a_{n_j}^{\max}$ in $n_j$ as the $i+1$--th ad (and thus $\Pi[i,j] = \Lambda_{i} \cdot q_{a^{\max}_{n_j}, n_j} \cdot  \hat{r}_{a^{\max}_{n_j}} + \Pi[i+1,j+1]$) or to leave node $n_j$ empty (and thus $\Pi[i,j] = \Pi[i,j+1]$). At the end of the execution, $\Phi[1,1]$ contains the optimal allocation of ads into the nodes of the path. The complexity of the algorithm is $O(|C|\cdot|N|)$.
%
%\vspace{-0.35cm}
%\vspace{-0.2cm}
\begin{algorithm}
\begin{algorithmic}[1]
\begin{scriptsize}
\FOR{all $i \in \{0, \ldots, |N|-1\}$} \label{st:lastclbeg}
	\STATE $\Pi[i,|N|] = \Lambda_{i} \cdot q_{a^{\max}_{n_{|N|}}, n_{|N|}} \cdot \hat{r}_{a^{\max}_{n_{|N|}}}$
	\STATE $\Phi[i,|N|] = \{(a^{\max}_{n_{|N|}}, n_{|N|})\}$ \label{st:lastclend} 
\ENDFOR
\STATE $j = |N|-1$\label{st:otherbeg}
\WHILE{$j \geq 1$}
	\FOR{all $i \in \{0, \ldots, j-1\}$}
		\IF {$\Pi[i,j+1] \geq \Lambda_{i} \cdot q_{a^{\max}_{n_j}, n_j} \cdot  \hat{r}_{a^{\max}_{n_j}} + \Pi[i+1,j+1]$}
			\STATE $\Pi[i,j] = \Pi[i,j+1]$ and $\Phi[i,j] = \Phi[i,j+1]$
		\ELSE
			\STATE $\Pi[i,j] = \Lambda_{i} \cdot q_{a^{\max}_{n_j}, n_j} \cdot  \hat{r}_{a^{\max}_{n_j}} + \Pi[i+1,j+1]$
			\STATE $\Phi[i,j] = \Phi[i+1,j+1] \cup \{(a^{\max}_{n_j}, n_j)\}$
		\ENDIF
	\ENDFOR
	\STATE $j=j-1$ \label{st:otherend}
\ENDWHILE
\RETURN{$\Phi[1,1]$}
\end{scriptsize}
\end{algorithmic}
\caption{}
\label{alg:maximal}
\vspace{-0.1cm}
\end{algorithm}
%\vspace{-0.2cm}
%
%
Algorithm~1 can be extended to find the optimal allocation even when some ads are maximal in more than one node. The idea of the algorithm is to enumerate all the possible ways to remove all the conflicts (as explained below) and then, for each combination, to compute the best allocations using Algorithm~1. First, choose an ad $a$ that is maximal in more than one node. To remove conflicts, for each node in which $a$ is maximal, the algorithm creates a new problem where $a$ is maximal only for that node and  is removed from the set of ads that can be displayed in the other conflicting nodes. This procedure removes some conflicts, but may generate new ones due ads becoming maximal in the nodes where $a$ has been removed. This algorithm iteratively proceeds until each node has a different maximal ad and, thus, we can apply Algorithm~1. Finally, the algorithm returns the advertising plan that maximises the $SW$ among the plans returned by all the executions of  Algorithm~1. 
In the worst case, the complexity  is $O\left(|N|^{\min\{|N|,|A|\}} |C| |N|\right)$ that is worse than the complexity of \textsf{OptimalSinglePath}. %More details, omitted here for reasons of space, can be found in~\cite{supplementalmaterial}.

\textbf{Economic mechanism}. We can have an AE, DSIC, IR, and WBB mechanism by resorting to the VCG mechanism with Clarke pivoting. Transfers $t_a$ can be easily found by using the algorithm for the allocation function $f_E$.
Formally, $t_a = SW(f_E(\hat{\bf{r}}_{-a})) - SW_{-a} (f_E(\hat{\bf{r}}))$, where $f_E(\hat{\bf{r}}_{-a})$ returns the optimal allocation when ad $a$ does not participate to the auction and $SW_{-a} (f_E(\hat{\bf{r}}))$ is the $SW$ of the optimal allocation when $a$ participates, but his contribution is not considered in the $SW$. The complexity of the mechanism is $\min\{|A|,$ $|N|\}$ times the complexity of the adopted $f_E$.

%\vspace{-0.2cm}
\subsection{Approximate Mechanisms}
\label{sec:approx}
%\vspace{-0.05cm}
\noindent Since the efficient mechanisms in the unrestricted domains  discussed above do not scale, it is important to consider approximate algorithms. Existing results show that 3AP does not admit any polynomial--time approximation scheme (PTAS), but it does admit a constant--ratio (the best one is $\frac{1}{2}$) approximation algorithms~\cite{Spieksma2000}. These approximation algorithms are based on the similarity between 3AP and the \emph{weighted $k$--set packing problem} (W$k$SPP) and the existence of approximation algorithms with ratio $O(\frac{1}{k})$ for this latter problem~\cite{arkin1998}. Specifically, any 3AP can be formulated as a W$k$SPP with $k=3$. However, our allocation problem cannot be formulated as W$k$SPP due to additional Constraints~(\ref{eq::ordered}) that cannot be formulated as set packing constraints of the form $\sum\cdot\leq1$. Thus, we cannot resort to such approximation algorithms. However, it is possible to design an \emph{ad hoc} polynomial--time approximation algorithm with constant approximation ratio w.r.t. both~$|N|$ and~$|A|$. We start by stating the following:
%, which allows us to find bounds from the optimal solution:
\vspace{-0.1cm}
\begin{proposition}
\label{le::delta}
Suppose we limit the total number of ads allocated, such that the continuation probability, $\Lambda_{c}$, of the last ad is at least $\delta$, i.e. $\forall n \in N: \Lambda_{c(\theta,n)} \geq \delta $. Then, the optimal social welfare given the reduced allocation space is at least $(1-\delta)$ the optimal social welfare when considering the set of all possible allocations~$\Theta$. 
% Given $\delta \in(0,1)$, the optimal solution of the single--path allocation problem within a subset of possible allocations, \ where the last displayed ad is reached with a cumulative continuation probability  is never worse than 
\end{proposition}
\vspace{-0.1cm}

\begin{proof}
Let OPT denote the social welfare of the optimal allocation and $M\subseteq N$ the set of nodes in which non--empty ads are allocated. With slight abuse of notation, we denote by $a_m$ the ad allocated in node $m$ and by $c_m$ the number of non--empty ads allocated in nodes preceding $m$. Divide $M$ into two subsets as follow: $M^+$ is the set of nodes $m$ with $\Lambda_{c_m}\geq \delta$ and $M^-$ is the set of nodes $m$ with $\Lambda_{c_m} < \delta$. We can write:

\begin{scriptsize}
\begin{align*}
OPT &= \sum_{m \in M} \Lambda_{c_m} \cdot \hat{r}_{a_m} \cdot q_{a_m,m} \\
&= \sum_{m \in M^+} \Lambda_{c_m}  \cdot \hat{r}_{a_m} \cdot q_{a_m,m} + \sum_{m \in M^-} \Lambda_{c_m}  \cdot \hat{r}_{a_m} \cdot q_{a_m,m}
\end{align*}
\end{scriptsize}

\noindent where $\sum_{m \in M^+} \Lambda_{c_m}  \cdot \hat{r}_{a_m} \cdot q_{a_m,m}$ is the social welfare of the truncated solution that allocate ads only to nodes $m \in M^+$.

Now consider the same allocation in which all the ads that are allocated to nodes $m \in M^+$ are removed. The resulting social welfare is $\sum_{m \in M^-} \Lambda'_{c_m}  \hat{r}_{a_m} q_{a_c,m}$, where $\Lambda'_{c_m}$ is the aggregated continuation probability of node $m$ when only $a_{m'}$ with $m' \in M^-$ are allocated.

Given the definition of $M^+$ and $M^-$, we know that $\frac{1}{\delta} \Lambda_{c_m} \leq \Lambda'_{c_m}, \forall m \in M^-$, and that:

\begin{scriptsize}
\begin{equation} \label{eq1} 
	\sum_{m \in M^-} \Lambda'_{c_m}  \cdot  \hat{r}_{a_m} \cdot  q_{a_m,m} \geq \frac{1}{\delta} \sum_{m \in M^-} \Lambda_{c_m}  \cdot  \hat{r}_{a_m} \cdot  q_{a_m,m}
\end{equation}
\end{scriptsize}

\noindent Moreover, by definition of OPT, we know that:

\begin{scriptsize}
\begin{equation}\label{eq2}
	OPT \geq \sum_{m \in M^-} \Lambda'_{c_m}  \cdot \hat{r}_{a_m} \cdot q_{a_m,m}
\end{equation}
\end{scriptsize}

\noindent Combining Equation~(\ref{eq1}) with Equation~(\ref{eq2}) we obtain  
that $\delta \cdot OPT \geq \sum_{j \in M^-} \Lambda_{c_m} \cdot \hat{r}_{a_m} \cdot q_{a_m,m}$, and thus:
\vspace{-0.2cm}

\begin{scriptsize}
\[ 
OPT - \sum_{j \in M^-} \Lambda_{c_m}  \cdot \hat{r}_{a_m} \cdot q_{a_m,m} \geq OPT - \delta \cdot OPT 
\]
\end{scriptsize}

\vspace{-0.2cm}
\noindent finally:
\vspace{-0.2cm}

\begin{scriptsize}
\[ 
\sum_{M  \in M^+} \Lambda_{c_m}  \cdot  \hat{r}_{a_m} \cdot  q_{a_m,m} \geq (1 - \delta) \cdot  OPT
\]
\end{scriptsize}

\vspace{-0.2cm}
\noindent This completes the proof of the proposition. %\hfill $\Box$
\end{proof}

We now present our approximate algorithm, $f_A$, which is a slight modification of the \textsf{OptimalSinglePath} algorithm.
% presented in the previous section.
The basic idea is that the exponential nature of the algorithm can be eliminated by fixing the maximum number of allocated non--empty ads to a given~$\overline{m}$. The algorithm generates all the possible combinations $B$ with $|B|\leq \overline{m}$ and then finds the optimal allocation for each combination $B$ by calling a 2AP--solving oracle.
\vspace{-0.1cm}
\begin{proposition}
Algorithm $f_A$ has a polynomial computational complexity $O(|N|^{\overline{m}} \cdot |A|^3)$ and  is an $(1- \prod_{i=1}^{\overline{m}-1} \lambda_i)$--approximation algorithm.
\end{proposition}
\vspace{-0.1cm}

\begin{proof}
Consider the partial permutations of $|N|$ elements in at most $\overline{m}$ positions. The number of partial permutations is $|N| + |N|\cdot (|N|-1) + \ldots + |N| \cdot (|N|-1) \cdots (|N| -\overline{m} + 1) = O(|N|^{\overline{m}})$. Since we are interested in combinations and not in permutations, we can overbound the number of combinations by $O(|N|^{\overline{m}})$. We can conclude that the computational complexity of the whole algorithm is  $O(|N|^{\overline{m}} \cdot |A|^3)$, since the complexity of the Hungarian algorithm is $O(|A|^3)$.%Consider the dispositions of $|N|$ elements in at most $\overline{m}$ positions. The number of possible dispositions are $|N| + |N|\cdot (|N|-1) + \ldots + |N| \cdot (|N|-1) \cdots (|N| -\overline{m} + 1) = O(|N|^{\overline{m}})$. Given that we are not interested in the order of the extracted elements, we would like to know the number of possible combinations of $|N|$ elements in at most $\overline{m}$ positions. But the combinations are less than the dispositions, thus $O(|N|^{\overline{m}})$,  and the complexity of the Hungarian algorithm is $O(|A|^3)$. We can conclude that the computational complexity of the whole algorithm is  $O(|N|^{\overline{m}} \cdot |A|^3)$.
This concludes the first part of the proof. Focus now on the approximation bound.

Given that $f_A$ allocates at most $\overline{m}$ ads, all the allocated non--empty ads have an aggregated continuation probability of $\Lambda_c \geq \prod_{i=1}^{\overline{m}-1} \lambda_i$. Thus, applying Proposition~\ref{le::delta} with $\delta = \prod_{i=1}^{\overline{m}-1} \lambda_i$, we can conclude that $f_A$ algorithm provides an approximation of $(1-\prod_{i=1}^{\overline{m}-1} \lambda_i)$ w.r.t. the optimal solution. %\hfill $\Box$
\end{proof}

It is worth noting that $f_A$ does not guarantee a constant approximation ratio given that  $\lambda_i$ can be arbitrarily close to 1 and, therefore, the bound can be arbitrarily close to 0. However, the approximation ratio does not depend on $N$ and~$A$ and therefore the algorithm scales to large instances. We remark that when $\lambda_i$ is close to 1, it would seem ``natural'' to approximate our allocation problem as a 2AP, by rounding $\lambda_i$ to 1. This new algorithm is denoted by $f_{A_2}$. However, we can state the following negative result.

%this algorithm, henceforth denoted by $f_{A_2}$, has an approximation ratio of $\prod_{i=1}^{|N|-1} \lambda_i$.

\vspace{-0.15cm}
\begin{proposition}
$f_{A_2}$ is an $\prod_{i=1}^{|N|-1} \lambda_i$--approximation algorithm, but is not monotone.
\end{proposition}
\vspace{-0.15cm}

\begin{proof} Let $SW(\theta)=\sum_{n \in N} \Lambda_{c(\theta, n)} q_{\theta(n), n} \hat{r}_{\theta(n)}$ and $\widetilde{SW}(\theta)=\sum_{n \in N} q_{\theta(n), n} \hat{r}_{\theta(n)}$. The allocations that maximise the two social welfares  are $f_E(\hat{\bf r}) = \theta^* = \arg \max_{\theta \in \Theta} SW(\theta)$ and $f_{A_2}(\hat{\bf r}) = \tilde{\theta} = \arg \max_{\theta \in \Theta} \widetilde{SW}(\theta)$. $c_{\max}$ is the number of allocated ads in $\tilde{\theta}$.

\begin{scriptsize}
\vspace{-0.5cm}
\begin{align*}
SW(\tilde{\theta}) & =\sum_{n \in N} \Lambda_{c(\tilde{\theta}, n)} q_{\tilde{\theta}(n), n} \hat{r}_{\tilde{\theta}(n)}  \geq  \Lambda_{c_{\max}-1} \sum_{n \in N}  q_{\tilde{\theta}(n), n} \hat{r}_{\tilde{\theta}(n)} \\
& \geq \Lambda_{c_{\max}-1} \sum_{n \in N}  q_{\theta^*(n), n} \hat{r}_{\theta^*(n)}\\ &\geq \Lambda_{c_{\max}-1} \sum_{n \in N}  \Lambda_{c(\theta^*, n)} q_{\theta^*(n), n} \hat{r}_{\theta^*(n)}\\
&= \Lambda_{c_{\max}-1} \cdot SW(\theta^*)
\end{align*}
\vspace{-0.5cm}
\end{scriptsize}

In the worst case, $c_{\max} = |N|$, thus the advertising plan produced by $f_{A_2}$ has an approximation ratio of $\prod_{i=1}^{|N|-1} \lambda_i$.

The non--monotonicity can be proved by counterexample.
Consider the case in Tab.~\ref{table::LTAP}(a) with $\lambda_i = 0.2\ \forall i \in \{1, \ldots, |N|-1\}$, $q_{a_1,n_1}=0.5$, and $q_{a_1,n_2}=1.0$. $f_{A_2}$ produces the allocation in which $a_1$ allocated in $n_1$ and $a_2$ in $n_2$; with this allocation $VTR_{a_1}=q_{a_1,n_1} = 0.5$.  Now consider the case in Tab.~\ref{table::LTAP}(b) where $r_{a_1}$ increases w.r.t. the previous case. $f_{A_2}$ produces the allocation in which $a_2$ is allocated in $n_1$ and $a_1$ in $n_2$; with this allocation  $VTR_{a_1}=\lambda_1 \cdot q_{a_1,n_2} = 0.2$. %\hfill $\Box$
\end{proof}

\textbf{Economic mechanism}. The adoption of $f_A$ as allocation function allows the definition of an incentive compatible mechanism in dominant strategies. 
\vspace{-0.1cm}
\begin{proposition}
$f_A$ is maximal in range.
\end{proposition}
\vspace{-0.1cm}

\begin{proof}
$f_A$ restricts the set of outcomes $\Theta$ to a subset $\Theta'\subseteq \Theta$ where only feasible allocations are those with at most $\overline{m}$ allocated ads. The restriction does not depend on the reports of the advertisers but only by information available to the mechanism. Given the reward declared by the advertisers, $f_A$ selects the allocation $\theta' = \arg\max_{\theta \in \Theta'} SW(\theta)$, thus it is maximal in range. %\hfill $\Box$
\end{proof}

As shown in~\cite{NisamRonen}, any allocation function that is maximal in range, if combined with VCG--based transfers with Clarke pivoting as
$t_a = SW(f_A(\hat{\bf{r}}_{-a})) - SW_{-a} (f_A(\hat{\bf{r}}))$,
leads to a DSIC mechanism. The mechanism satisfies also IR and WBB; the proof is easy by definition of VCG--based transfers.

%\vspace{-0.15cm}
\section{Multi--Path Case} \label{sec:multipath}
\noindent In this section, we extend the results previously discussed to the general case with multiple paths.

%\vspace{-0.1cm}
\subsection{Efficient Mechanism}

We focus only on the allocation function, referred to as $f_{EM}$ (since the VCG transfer with Clarke pivot can again be used to obtain a DSIC mechanism). We can formulate the problem of finding the optimal allocation as an integer linear program by extending the single--path formulation as follows:

%\vspace{-0.25cm}
\begin{scriptsize}
\begin{equation*}
\max \sum_{a \in A} \sum_{p \in P_v} \sum_{c \in C_p} \sum_{n \in N_p} \omega_p \cdot \Lambda_c \cdot \hat{r}_{a} \cdot q_{a,n} \cdot x_{a,n,c,p} 
\end{equation*}
\vspace{-0.25cm}
\begin{align}
\sum_{n \in N_p} \sum_{c \in C_p} x_{a,n,c,p}  \leq 1  &&  \forall a \in A,  p \in P_v \label{eq::onetimemulti}\\
\sum_{a \in A} \sum_{c \in C_p} x_{a,n,c,p}  \leq 1   &&  \forall p \in P_v, n \in N_p \label{eq::oneslotmulti}\\
\sum_{a \in A} \sum_{n \in N_p} x_{a,n,c,p}  \leq  1 &&  \forall c \in C_p, p \in P_v \label{eq::onepositionmulti}
\end{align}
\vspace{-0.5cm}
\begin{align}
\sum_{a \in A} x_{a,n,c,p}  - \sum_{a \in A} \sum_{\underset{n' < n}{n' \in N_p:}} x_{a,n',c-1,p} \leq 0 &\begin{split}   \forall  p \in P_v,n \in N_p,\\ c \in C_p \setminus \{0\} \end{split} \label{eq::orderedmulti}
\end{align}
\vspace{-0.5cm}
\begin{align}
x_{a,n,c,p} - x_{a,n,c,p'} = 0   & \begin{split}\forall p,p' \in P_v, n \in p \cap p',\\ a \in A,  c \in C_p \end{split}\label{eq::vincolomulti}\\
x_{a,n,c,p} \in \{0,1\}   & \begin{split}\forall a \in A,  n \in N,  \\  p \in P_v,c \in C_p\end{split} \label{eq::integermulti}
\end{align}
\end{scriptsize}

%\vspace{-0.25cm}
\noindent where $N_p$ and $C_p$ depend on the specific path $p$. Basically, the variables for the single--path case, $x_{a,n,c}$, are replicated for each path $p$, i.e., $x_{a,n,c,p}$. Each path $p$ must satisfy the same constraints we have in the single--path case and, in addition, Constraints~(\ref{eq::vincolomulti}) force nodes that are shared by multiple paths to be assigned to the same ad. The objective function maximizes the (expected) social welfare.
%We focus only on the allocation function, referred to as $f_{EM}$ (since the VCG mechanism can again be used to obtain DSIC). We can formulate the problem of finding the optimal allocation as an integer linear program by extending the single--path formulation.
%Basically, the variables for the single--path case, $x_{a,n,c}$, are replicated for each path $p$, i.e., $x_{a,n,c,p}$. Each path $p$ must satisfy the same constraints we have in the single--path case. In addition, the constraints that force nodes that are shared by multiple paths to be assigned to the same ad must be satisfied. Formally, $x_{a,n,c,p} - x_{a,n,c,p'} = 0$ $ \forall p,p' \in P_v,  a \in A,  c \in C_p, n \in p \cap p'$.
%The objective function maximizes the (expected) social welfare. 
We notice that differently from the single--path case, even when $\lambda=1$, our problem can no longer be formulated as a 2AP (it is a variation of the 2AP with additional constraints whose continuous relaxation admits non--integer solutions). In this case, we use the classical branch--and--bound algorithm whose complexity is~$O(|A|^{|N|})$.

\textbf{Allocation function algorithm in restricted domain}. As in the case of single--path, we can identify a restricted domain where $f_{EM}$ is computationally easy. For the single--path case we have shown that, when the nodes of the path have different maximal ads, the problem becomes easy. In the multi--path environment we can state something stronger: when, for each path $p$, all the nodes belonging  to $N_p$ have different maximal ads, $f_{EM}$ is computationally easy. Thus, we allow an ad to be maximal in multiple nodes, as long as these nodes belong to different paths.

An optimal algorithm for this restricted domain is given by Algorithm~\ref{alg:maximalmulti}, which extends Algorithm~\ref{alg:maximal} to the multi--path case. To this end, we need to define two additional functions: $s: N \rightarrow \mathcal{P}(N)$, which returns, for any node $n \in N$, $s(n)$, the set of children nodes of $n$ in the multi--path tree; and $l: N \rightarrow \mathbb{N}$, which returns, for any node $n \in N$, $l(n)$, the number of nodes on the path from the root of the tree, $n_1$, to node $n$ (including the root $n_1$ and $n$).

Algorithm~\ref{alg:maximalmulti} is based on a recursive procedure and proceeds as follows. The base case is reached when the parameter of the algorithm is a leaf node. Then, the algorithm builds the optimal advertising plan from the subpaths generated starting from the leaf nodes and backtracking until the root node $n_1$ is reached. Each call to the algorithm $f_{MP}(n)$ requires the allocation of two vectors $\Phi_n$ and $\Pi_n$ with size $l(n)$. $\Phi_n[i]$ and $\Pi_n[i]$ are the optimal allocation and its value respectively, in the subtree with root $n$ when the first displayed ad in the subtree will be the $i+1$--th in the tree allocation.

%For reasons of space we leave the detailed explanation of Algorithm~\ref{alg:maximalmulti}. The complexity of the algorithm is $O(|P_v|\cdot |C|\cdot|N|)$.
%\vspace{-0.3cm}
\begin{algorithm}[!h]
\begin{algorithmic}[1]
\begin{scriptsize}
\IF{$s(n)=\emptyset$}
	\FOR{all $i \in \{0,\ldots, l(n)-1\}$} \label{st:basecasebeg}
		\STATE $\Pi_n[i] = \alpha_n \cdot \Lambda_{i} \cdot q_{a^{\max}_{n}, n} \cdot  \hat{r}_{a^{\max}_{n}}$ and $\Phi_n[i] = \{(a^{\max}_{n},n)\}$ \label{st:basecaseend}
	\ENDFOR
\ELSE
	\STATE $\Pi_n[\cdot] = 0$ and $\Phi_n[\cdot] = \emptyset$ \label{st:recbeg}
	\FOR{all $n' \in s(n)$} \label{st:chbeg}
		\STATE $\Pi_{n'}, \Phi_{n'} = f_{MP}(n')$ \label{st:chend}
 	\ENDFOR 
	\FOR{all $i \in \{0,\ldots, l(n)-1\}$} \label{st:allocbeg}
		\IF {$\sum_{n' \in s(n)} \Pi_{n'}[i] \geq \alpha_n \cdot \Lambda_{i} \cdot q_{a^{\max}_{n}, n} \cdot \hat{r}_{a^{\max}_{n}} + \sum_{n' \in N} \Pi_{n'}[i+1]$}
			\STATE $\Pi_n[i] = \sum_{n' \in s(n)} \Pi_{n'}[i]$ \label{st:noallb} and $\Phi_n[i] = \cup_{n' \in s(n)} \Phi_{n'}[i]$ \label{st:noalle}
		\ELSE
			\STATE $\Pi_n[i] = \alpha_n \cdot \Lambda_{i} \cdot q_{a^{\max}_{n}, n} \cdot  \hat{r}_{a^{\max}_{n}} + \sum_{n' \in N} \Pi_{n'}[i+1]$ \label{st:yesallb}
			\STATE $\Phi_n[i] = \cup_{n' \in s(n)}\phi_{n'}[i+1] \cup \{(a^{\max}_{n}, n)\}$ \label{st:yesalle} \label{st:recend}
		\ENDIF
	\ENDFOR
\ENDIF
\RETURN{$\Pi_n, \Phi_n$}
\end{scriptsize}
\end{algorithmic}
\caption{$f_{MP}(n)$}
\label{alg:maximalmulti}
\end{algorithm}

Algorithm~\ref{alg:maximalmulti} starts its execution by being called with $f_{MP}(n_1)$, where $n_1$ is the root node of the multi--path tree. Steps~(\ref{st:basecasebeg}--\ref{st:basecaseend}) deal with the base case. If a leaf node $n$ has been reached, $\forall i$ $\Pi_n[i]$ is filled with the value provided by the maximal ad $a^{\max}_n$ in node $n$ when it will be the $i+1$--th displayed ad on its path, i.e. $\Pi_n[i] = \alpha_n \cdot \Lambda_{i} \cdot q_{a^{\max}_{n}, n} \cdot  \hat{r}_{a^{\max}_{n}}$. Then, the algorithm stores in $\Phi_n[i]$, $\forall i$, the fact that  $a_n^{\max}$ has been allocated in $n$. Indeed, in any optimal allocation an ad is allocated in the last node of each path. Steps~(\ref{st:recbeg}--\ref{st:recend}) deal with the recursion step. The algorithm, at Step~\ref{st:chend}, recursively call $f_{MP}(n')$ for all the children nodes $n'$ of $n$. Finally, the algorithm decides whether it is optimal to leave the node non--filled, Step~\ref{st:noallb}, or  to allocate $a_n^{\max}$ in $n$, Steps~(\ref{st:yesallb}--\ref{st:yesalle}). These last operations are similar to the one of Algorithm~\ref{alg:maximal}, except that, for node $n$, the value provided by the ad allocated has to be weighted by the probability the node is reached during the paths ($\alpha_n$). At the end of the execution of Algorithm~\ref{alg:maximalmulti}, $\Phi_{n_1}[1]$ stores the optimal advertising plan. The complexity of the algorithm is $O(|P_v|\cdot |C|\cdot|N|)$.

%\vspace{-0.6cm}

%%%%%%%%%%%%%
%%%%%%%%%%%%%
\subsection{Approximate Mechanism}
\noindent We now consider an approximate mechanism for the multi-path setting, and show that it is possible to provide a maximal--in--range approximation algorithm $f_{AM}$ with approximation ratio that is constant w.r.t. $|N|$ and $|A|$, but decreases with $|P_v|$. 
Let $SW_p(\theta) = \sum_{n \in N_p} \alpha_n \cdot \Lambda_{c(\theta, n)} \cdot q_{\theta(n), n} \cdot \hat{r}_{\theta(n)}$ denote the social welfare for a single path in the tree, and define $\theta^* = \arg\max_{\theta \in \Theta} SW(\theta)$ and $\theta^*_p = \arg\max_{\theta \in \Theta} SW_p(\theta)$.
Given this, the following holds:
\vspace{-0.05cm}
\begin{proposition}
The value $\max_{p\in P_v}\{SW_p (\theta^*_p)\}$ is never worse than $\frac{1}{|P_v|}$ of the optimal allocation for the entire tree.
\end{proposition}
\vspace{-0.05cm}

\begin{proof}
Given the definition before we know that $SW_p(\theta^*_p) \geq SW_p(\theta^*), \forall p \in P_v$, and that $\sum_{p \in P_v} SW_p(\theta^*) \geq SW(\theta^*) = OPT$. Thus, since $\sum_{p \in P_v}SW_p(\theta^*_p) \geq OPT$, we can state that $\max_{p\in P_v}\{SW_p (\theta^*_p)\} \cdot |P_v| \geq OPT$ and therefore $\frac{1}{|P_v|} \leq$\\ $\frac{\max_{p\in P_v}\{SW_p (\theta^*_p)\}}{OPT}$. This completes the proof. %\hfill $\Box$
%Easily, we can decompose the social welfare $SW$ in the multi--path case as the sum of the contributions due to the single paths as $SW=\sum_{p \in P_v} SW_p$. Thus, the ratio $\frac{\max_{p\in P_v}\{SW^*_p\}}{SW^*}\leq\frac{\max_{p\in P_v}\{SW^*_p\}}{\sum_pSW_p^*}\leq \frac{1}{|P_v|}$.This completes the proof of the proposition.
\end{proof}

By using this proposition, we can provide a simple approximation algorithm that computes the best allocation $\theta^*_p, \forall p \in P_v$ and
selects the allocation of the path with the maximum $SW^*_p(\theta^*_p)$, obtaining a bound of $\frac{1}{|P_v|}$. However, this algorithm requires exponential time, which is the same as finding the optimal allocation of the single--path problem. By approximating this latter problem as described in Section~3.2, we obtain a polynomial--time approximation algorithm with bound $\frac{1-\prod_{i=1}^{\overline{m}-1} \lambda_i}{|P_v|}$.
It is easy to see that the algorithm is maximal in range as in the single--path case, and therefore it is possible to design a DSIC, WBB, IR, VCG--based mechanism with Clarke pivoting.

%\vspace{-0.2cm}
\section{Experimental Evaluation} \label{sec:experimentalevaluation}
%\vspace{-0.1cm}
\noindent In our experiments we compare the run time and the quality of the solutions obtained using  the above algorithms.

\vspace{-0.05cm}
\textbf{Instance generation}. We represent the experimental environment by a 10$\times$10 grid map in which each cell corresponds to a vertex of graph $G$. We associate each advertiser $a$ with a cell, $s_a$, in which we place the shop of $a$. The reward $r_a$ is uniformly drawn from $[0,100]$. To generate paths, we randomly select a starting vertex $v_s$ and, from $v_s$, we build the paths moving randomly to the adjacent (horizontally and vertically) cells until the desired length of the path is reached. The quality $q_{a,n}$ is uniformly drawn from $[0,1]$ if $n=s_a$, and it is $\max\{0,q_{a,s_a} - d_a\cdot dist(s_a,v)\}$ if $n=v\neq s_a$, where $d_a$ is a coefficient uniformly drawn from $[0,1]$ and $dist(s_a,v)$ is the Manhattan distance between $s_a$ and $v$ (normalized w.r.t. the maximum Manhattan distance among two cells in the grid map). The basic idea is that the quality linearly decreases as the distance between the current node and $s_a$ increases, and $d_a$ gives the decreasing speed. We assume a constant continuation probability $\lambda_i = \lambda \ \forall i \in \{1, \ldots, |N|\}$. We generate $50$ instances for each of the following configurations: $\lambda = 0.5$ and $N \in \{10,20,30,40,50\}$, and $\lambda = 0.8$ and $N \in\{ 10,20,30\}$.  In all instances $|A|=30$. For our mathematical programming formulations we use AMPL as modeling language and CPLEX 11.0.1 to solve them. The experiments were conducted on an Unix computer with $2.33$GHz CPU, $16$Gb RAM, and kernel 2.6.32-45. 

\begin{figure}
\centering
\begin{minipage}{15cm}
%\hspace{-0.3cm}
\includegraphics[scale=0.3]{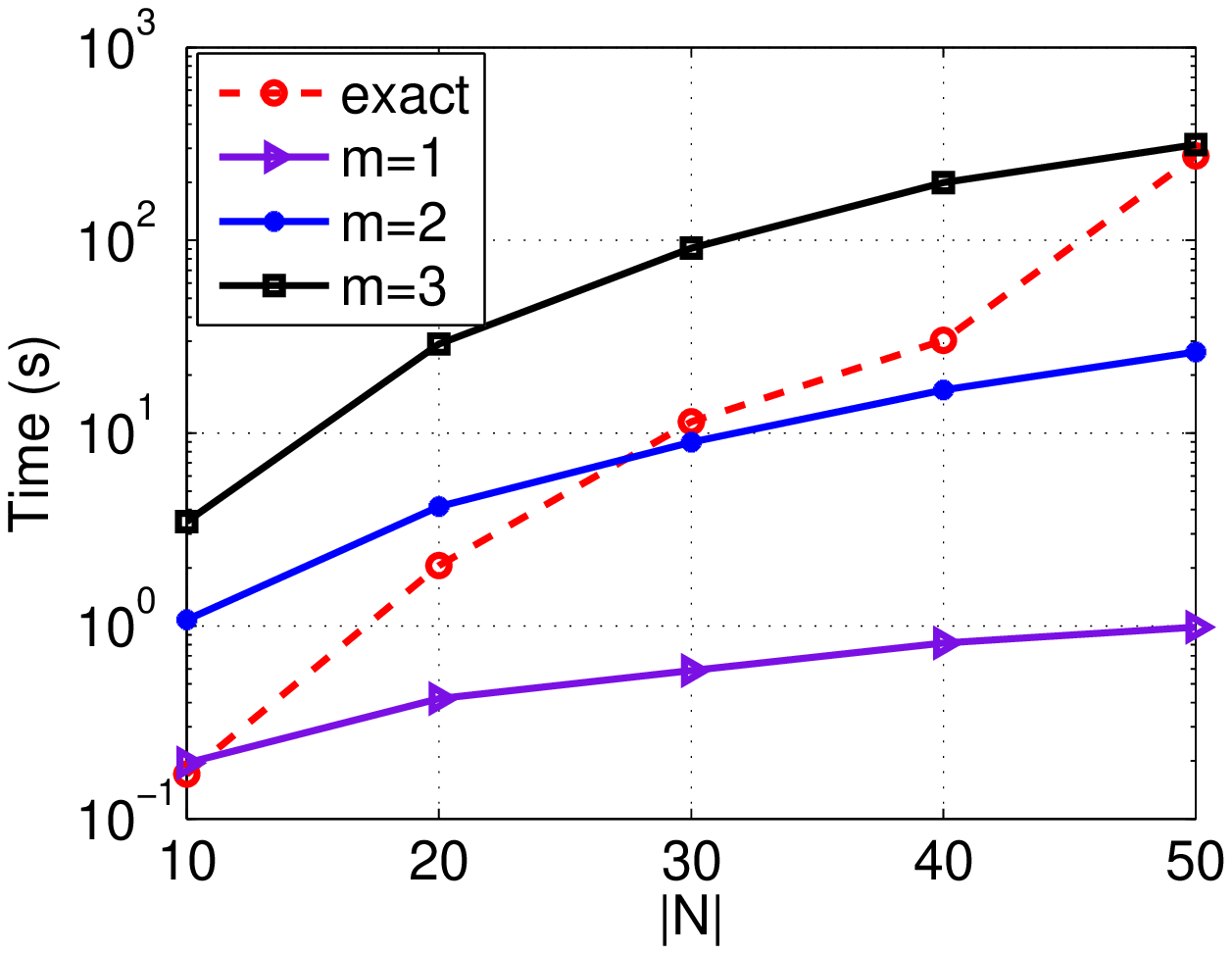}
\includegraphics[scale=0.3]{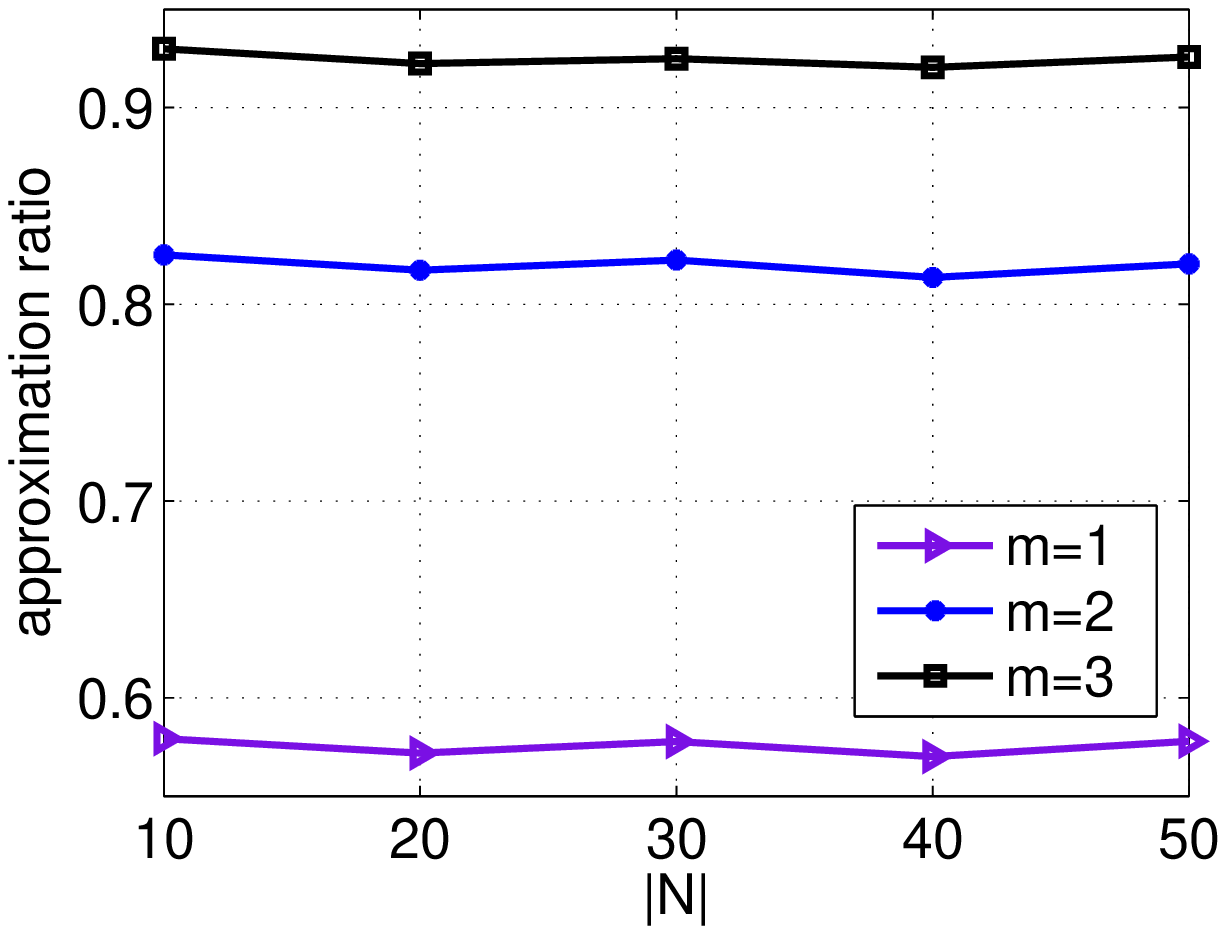}
\end{minipage}
\begin{minipage}{15cm}
%\hspace{-0.3cm}
\includegraphics[scale=0.3]{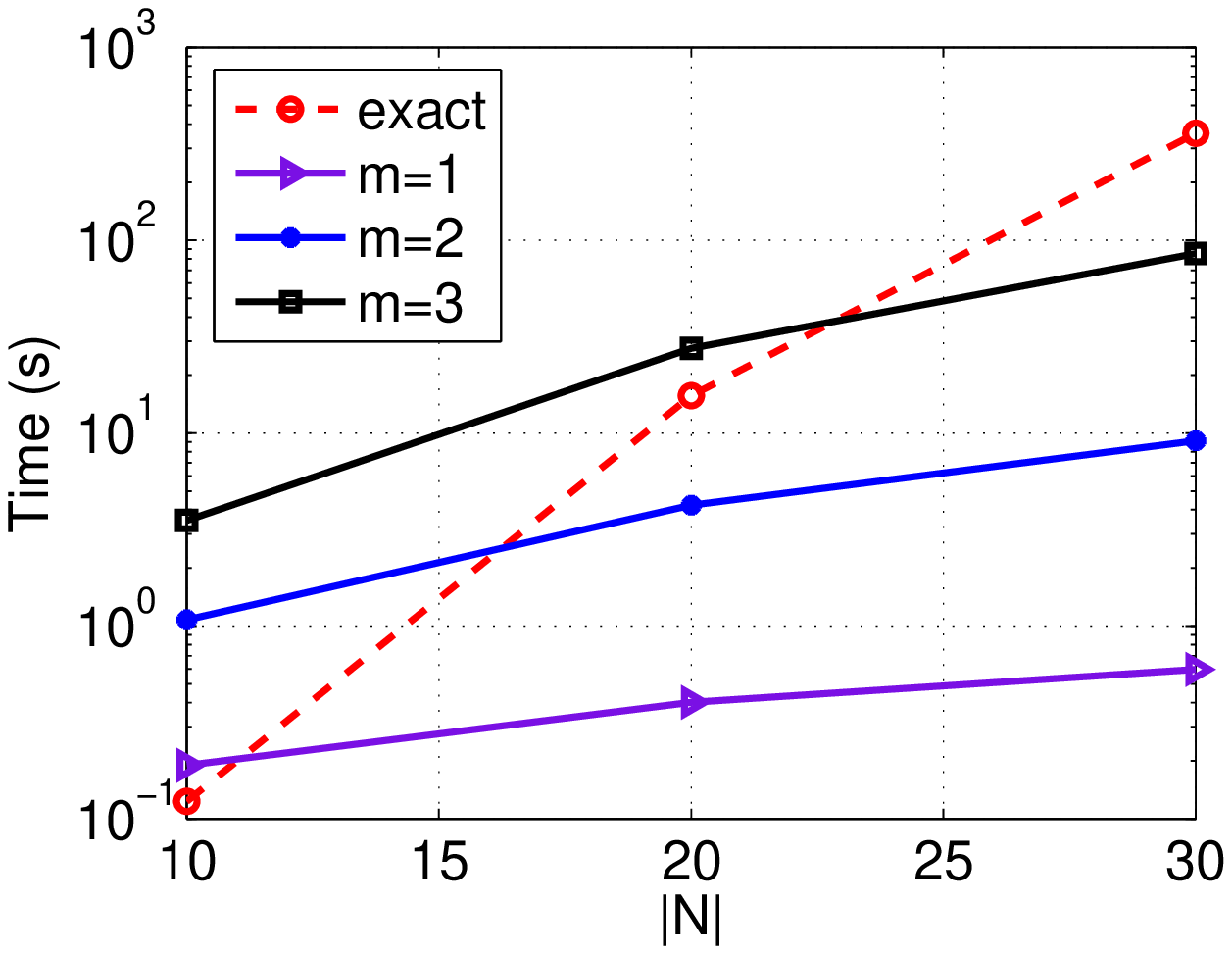}
\includegraphics[scale=0.3]{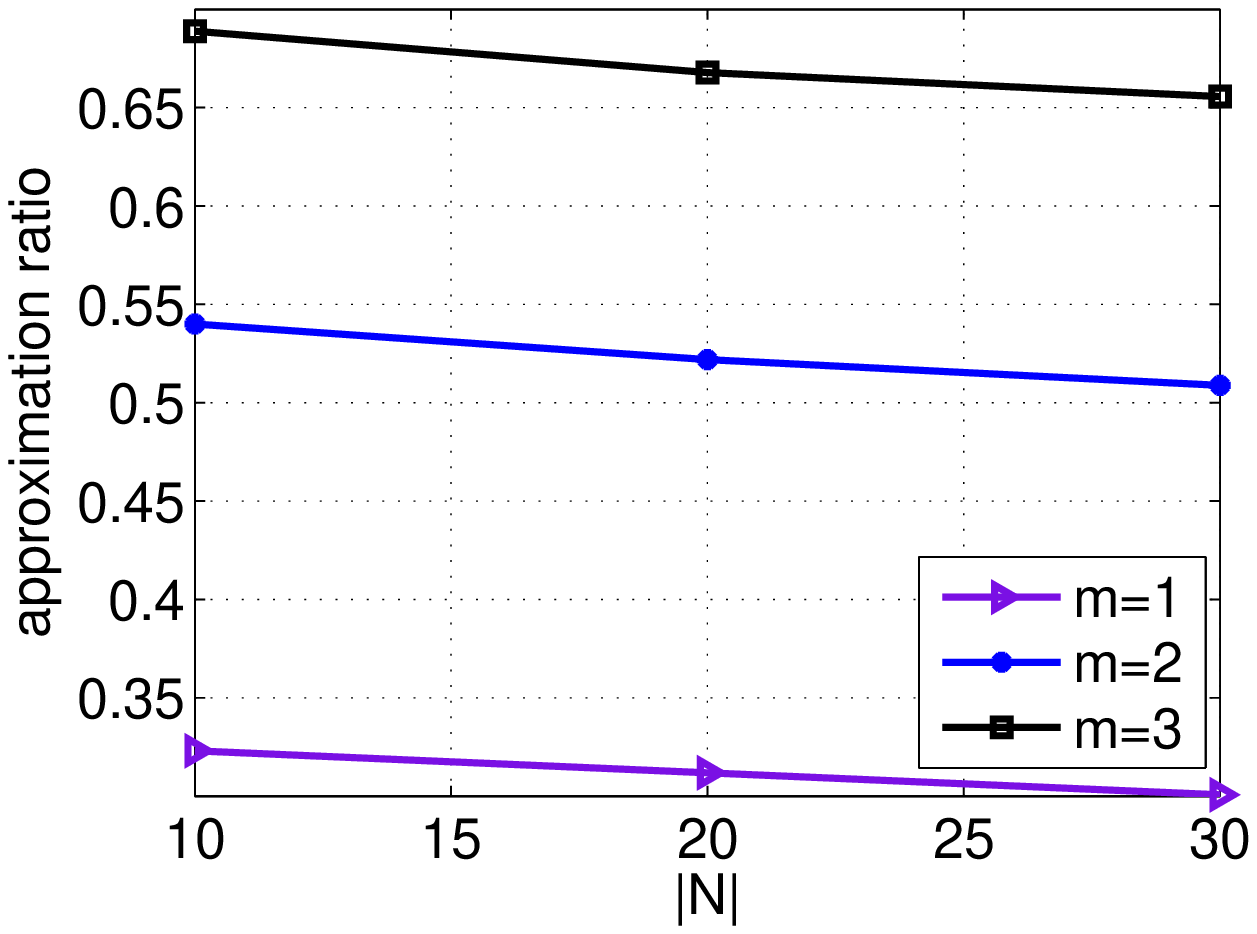}
\end{minipage}
\vspace{-0.5cm}
\caption{Average run time (left) and approximation ratio (right) as $|N|$ varies. $\lambda = 0.5$ (top) and $\lambda = 0.8$ (bottom).}
%\vspace{-0.45cm}
\label{fig::exp}
\end{figure}

\textbf{Single--path results}. We ran $f_E$ (specifically, the \textsf{OptimalSinglePath} implementation) and $f_A$ with $\overline{m} \in \{1,2,3\}$. The results are depicted in Fig.~\ref{fig::exp}. The average run time (left) and the average approximation ratio (AAR) obtained with different $\overline{m}$ (right) are plotted as $|N|$ varies. The two top plots  are with $\lambda =0.5$, while the two bottom plots are with $\lambda=0.8$. We observe that the run time of $f_E$ strictly depends on $\lambda$: the larger $\lambda$, the longer the run time. This is because, in the optimal allocation, the number of allocated ads increases as $\lambda$ increases (7 with $\lambda=0.5$ and 16 with $\lambda=0.8$), requiring a larger number of possible allocations to be considered. With $\lambda = 0.5$, $f_E$ can be used in practice to solve instances with a large number of nodes (up to $50$) within $10^3$~s, while, with $\lambda=0.8$, $f_E$ cannot be used for $|N|>30$ (we found instances that were not solved even after 10 hours). Instead, the run time of $f_{A}$ is constant in $\lambda$, and, differently from the worst--case complexity, run time is sub linear in $|N|$. On the other hand, the AAR, as theoretically expected, decreases as $\lambda$ increases. However, $f_A$ largely satisfies the theoretical bound, e.g., with $\lambda=0.5$ and $\overline{m}=2$, the theoretical bound is $0.5$, while we experimentally observed an AAR of $0.83$. %In addition, the AAR is, as the theoretical bound, independent of $|N|$.

\begin{figure}
%\begin{center}
\centering
\begin{minipage}{15cm}
%\hspace{-0.3cm}
\includegraphics[scale=0.3]{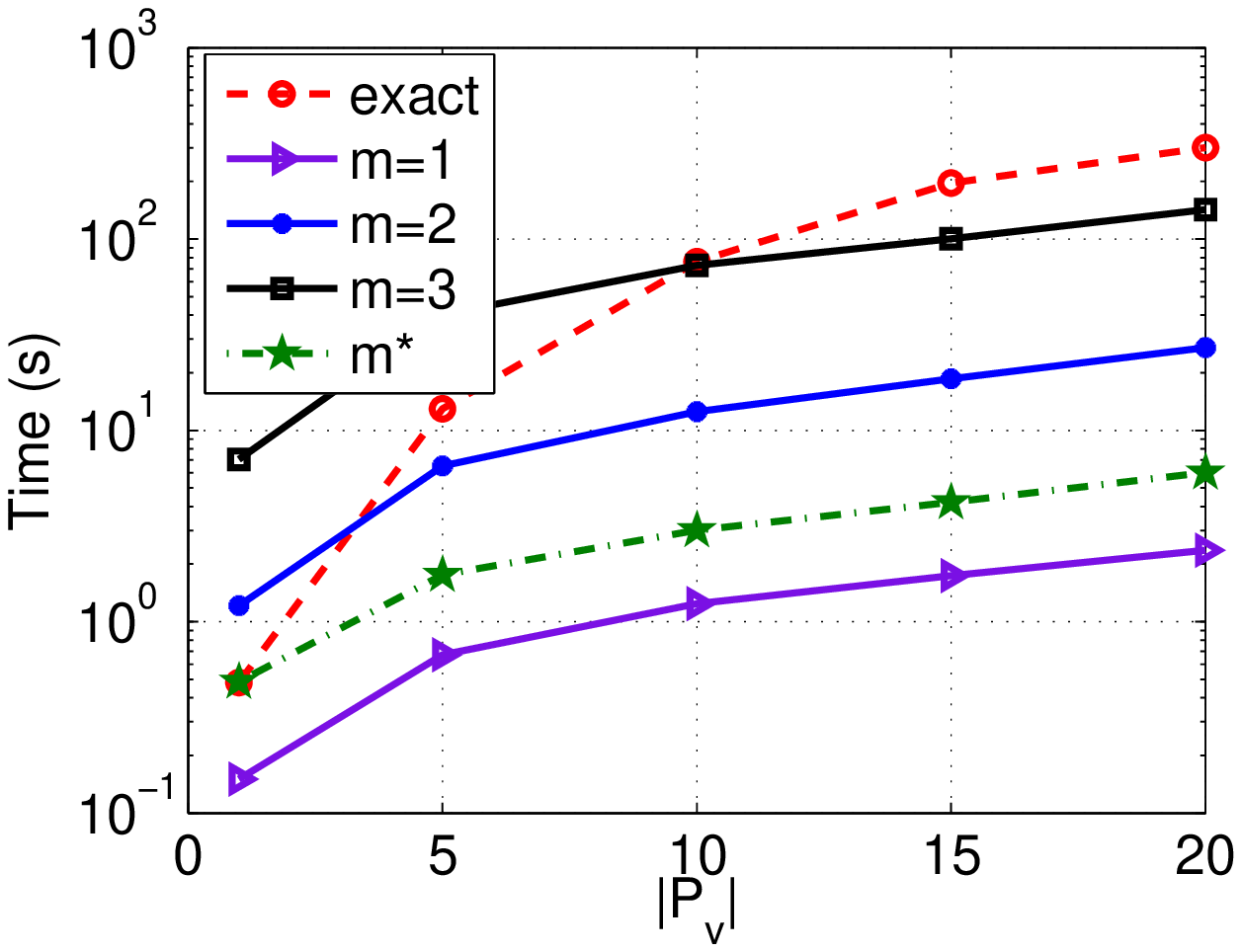}
\includegraphics[scale=0.3]{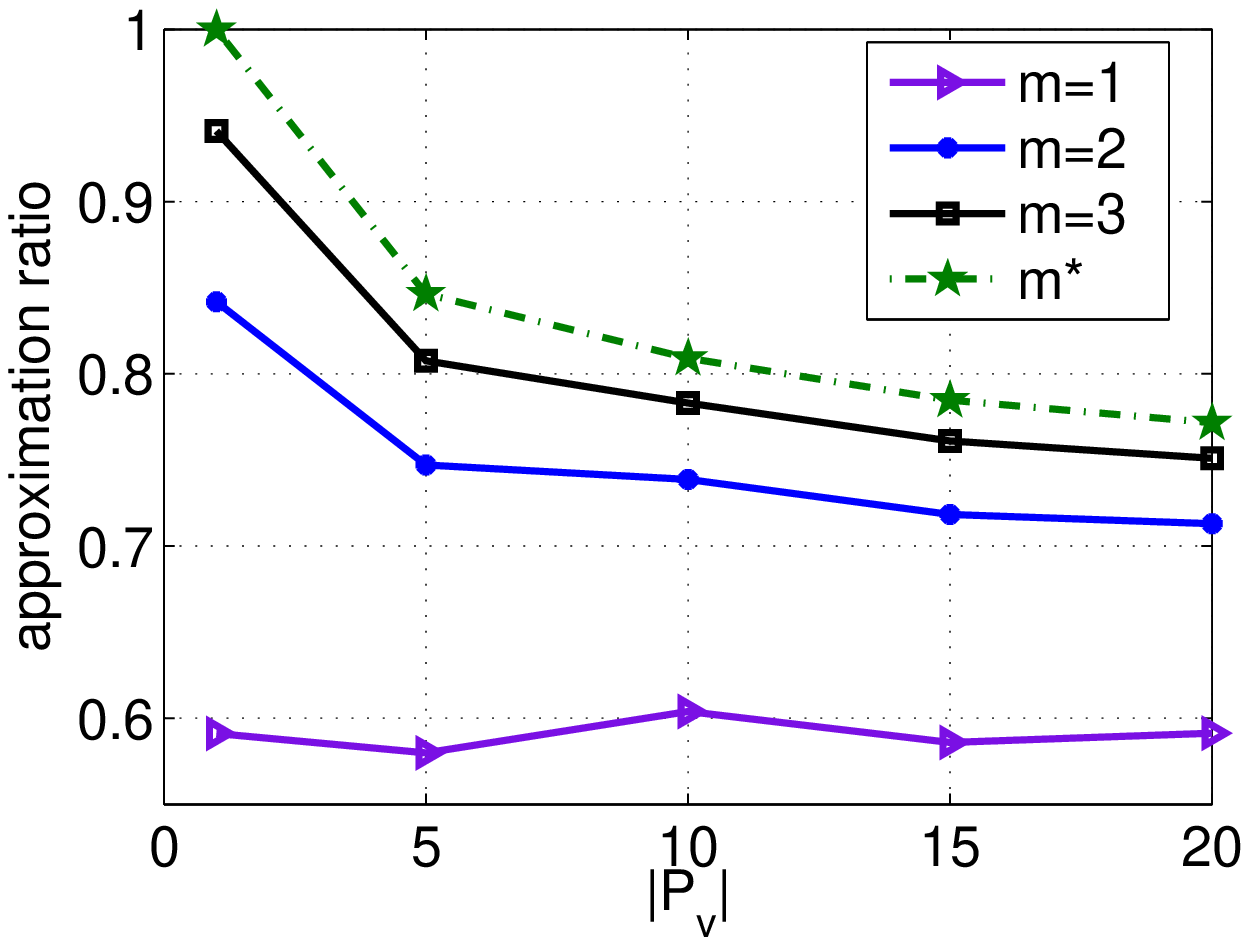}
\end{minipage}
%\begin{minipage}{15cm}
%\hspace{-0.3cm}
%\includegraphics[scale=0.3]{figures/time05tree}
%\includegraphics[scale=0.3]{figures/value05tree}
%\end{minipage}
\vspace{-0.5cm}
\caption{Average run time (left) and approximation ratio (right) as $|P_v|$ varies with $\lambda = 0.5$ in the multi--path case.}
%\vspace{-0.7cm}
\label{fig::expMulti}
%\end{center}
\end{figure}

\textbf{Multi--path results}. We ran $f_{EM}$ and $f_{AM}$ with $\overline{m} \in \{1,2,3\}$. The results are depicted in Fig.~\ref{fig::expMulti}. By $m^*$ we denote a variation of $f_{AM}$ in which we adopt $f_E$ to find the optimal solution $\theta^*_p$ on the single path $p$ (used because, as discussed above, the run time of $f_E$ is tractable for 20 nodes or less). The figures show, for $\lambda=0.5$, the average run time (left) and the AAR obtained with different values of $\overline{m}$ (right) as $|P_v|$ varies, while the length of each path is uniformly drawn from $\{1,\ldots,20\}$. With $|P_v|=15$ and $20$, we interrupted the execution of $f_{EM}$ in 2 instances due to the set time limit (1200~s); with $\lambda=0.8$, the number of interrupted executions is 11 when $|P_v|=20$. Thus, $f_{EM}$ can be used in practice with instances with no more than $20$ paths and a small $\lambda$. We experimentally observed that AARs are much better than the theoretical bounds. In particular, the theoretical bound decreases as $\frac{1}{|P_v|}$, instead, experimentally, the ratios seem to converge to values~$\geq 0.6$ as $|P_v|$ increases. Also, $m^*$ provides the best performance in terms of trade--off between run time and AAR.

%\begin{table}[!h]
%\begin{tabular}{cc}
%
%\includegraphics[width=.26\textwidth]{figures/time05single} & \includegraphics[width=.26\textwidth]{figures/value05single}\\
%
%
%\end{tabular}
%\end{table}

%\vspace{-0.3cm}
\section{Conclusions and Future Works} \label{sec:conclusions}
%\vspace{-0.1cm}

In this paper, we introduced, for the first time, an economic model for mobile geo--location advertising.
We designed a user mobility model whereby the user moves along one of several paths and we designed some incentive compatible mechanisms: exact with exponential time, exact with polynomial time for a significant restricted set of instances, and 
%We focused on the allocation problem, and showed that the problem of finding the optimal advertising plan can be formulated as a variation of the assignment problem. In addition to optimal algorithms, we developed polynomial--time approximate algorithms with theoretical bounds.
approximate with theoretical bounds and polynomial--time.
Finally, we experimentally evaluated our algorithms in terms of the trade--off between sub--optimality of the allocation and compute time showing that in the single--path case the optimal solution can be found for large instances and that the average--case approximations we found are significantly better than the worst--case theoretical bound. With multi--path cases, finding the optimal allocation quickly becomes intractable, and approximation algorithms are necessary.

In future work, we aim to prove the $\mathcal{NP}$--hardness of our allocation problem and to design more efficient (even non--monotone) approximation algorithms. %Furthermore, we aim to explore machine learning techniques for the estimation of the user model parameters and to evaluate the proposed model in real--world applications.

%\vspace{-0.1cm}


\bibliography{citations}
\bibliographystyle{aaai}

\end{document}